\documentclass[preprint,12pt]{elsarticle}

\usepackage{amssymb}
\usepackage{amsmath}

\journal{Computers And Fluids}

\begin{document}

\begin{frontmatter}



\title{A Deep Learning Approach For Epistemic Uncertainty Quantification Of Turbulent Flow Simulations}

\author[inst1]{Minghan Chu}

\affiliation[inst1]{organization={Mechanics and Materials Engineering Department},
            addressline={130 Stuart Street}, 
            city={Kingston},
            postcode={K7L 3N6}, 
            state={ON},
            country={Canada}}

\author[inst2]{Weicheng Qian}

\affiliation[inst2]{organization={Department of Computer Science, University of Saskatchewan},
            addressline={176 Thorvaldson Bldg, 110 Science Place}, 
            city={Saskatoon},
            postcode={S7N 5C9}, 
            state={SK},
            country={Canada}}

\begin{abstract}
Simulations of complex turbulent flow are part and parcel of the engineering design process. Eddy viscosity based turbulence models represent the workhorse for these simulations. The underlying simplifications in eddy viscosity models make them computationally inexpensive but also introduce structural uncertainties in their predictions. Currently the Eigenspace Perturbation Method is the only approach to predict these uncertainties. Due to its purely physics based nature this method often leads to unrealistically large uncertainty bounds that lead to exceedingly conservative designs. We use a Deep Learning based approach to address this issue. We control the perturbations using trained deep learning models that predict how much to perturb the modeled Reynolds stresses. This is executed using a Convolutional Neural Network that learns the difference between eddy viscosity based model predictions and high fidelity data as a mapping of flow features. We show that this approach leads to improvements over the Eigenspace Perturbation Method.
\end{abstract}




\end{frontmatter}


\section{Background And Motivation}
Fluid turbulence is an important problem in a variety of problems of engineering design. In spite of research there is no analytical theory that can predict the evolution of complex, real-life turbulent flows. Engineering design studies have to use turbulence models for simulations. These turbulence models are simplified constitutive equations that relate quantities of interest that are challenging to compute to known and easily computable quantities. 

There are different turbulence modeling approaches available for engineering design ranging from Large Eddy Simulations (LES) \cite{lesieur1996new, lesieur2005large} and Reynolds Stress Modeling (RSM) \cite{speziale1991analytical, launder1975progress, mishra2014realizability, speziale1991modelling} to Eddy Viscosity Models (EVM) \cite{craft1996development, gatski2002linear, shih1995new, kraichnan1976eddy}. These approaches differ in the proportion of turbulence that is resolved and that which is modeled at different levels of fidelity \cite{speziale1991analytical, pope2001turbulent}. This also leads to a gradation in the computational costs, where LES represents the most expensive approach from a computational viewpoint. Eddy Viscosity Models provide reasonably accurate predictions for many complex flows at a low computational cost. Eddy viscosity based turbulence models represent the workhorse for flow simulations for engineering design.

Eddy viscosity models use large simplifications in their formulation to retain computation economy and robustness. These include the gradient diffusion hypothesis \cite{da2007analysis} and the turbulent viscosity hypotheses (TVH) \cite{schmitt2007boussinesq}. These simplifications delimit the extent to which eddy viscosity models can reflect turbulent physics of complex flows such as those with rotational effects, streamline curvature, flow separation, etc. This leads to epistemic uncertainty in their predictions. 

Epistemic uncertainties in turbulent flow simulations are produced by limited understanding of turbulence physics and the limitations in incorporating physics accurately in turbulence models \cite{smith2013uncertainty}. Epistemic uncertainties in turbulence models occur due to many reasons including limited understanding of the turbulence physics \cite{duraisamy2019turbulence}, simplifications incorporated to make the turbulence model applicable for engineering workflows \cite{oliver2011bayesian, mishra2016sensitivity}, simplifications made to make the turbulence model computationally inexpensive \cite{alonso2017scalable}, paucity of data to tune the model \cite{kato2013approach}, etc. This epistemic uncertainty can be structural uncertainty due to the structure of the turbulence model expressions \cite{dow2011quantification} and parameter uncertainty due to the inferred values of the closure coefficients in the model expressions. The structural uncertainty is often the dominant source of errors and uncertainties in turbulent flow simulations for complex real life flows of engineering interest. In the iterative process of engineering design where hundreds of intermediate designs are sequentially evaluated using CFD simulations to optimize the final design these errors and uncertainties can have a substantial impact leading to significantly sub-optimal designs. So having reliable turbulence model structural uncertainty estimates is essential for reliable engineering designs. Novel engineering design approaches like robust design and reliability based design fall under the umbrella of the Design Under Uncertainty (DUU) approach \cite{yao2011review, padula2006aerospace}. The success of this approach depends on the quality of the uncertainty estimates provided to the approach. So, reliable and calibrated uncertainty estimates of turbulence model structural uncertainty are essential for engineering design.

At the present moment the Eigenspace Perturbation Method (EPM) \cite{iaccarino2017eigenspace} is the only data-free approach for turbulence model structural uncertainty quantification. The EPM uses sequential perturbations to the eigenvalues, eigenvectors and the amplitude of the modeled Reynolds stress tensor. Propagating CFD simulations via these perturbed Reynolds stresses leads to a discrete set of predictions. The union over these predictions is used as a measure of the structural uncertainty due to the turbulence models. In the past the EPM has been applied across different fields of engineering including the design of urban canopies\cite{gorle2019epistemic}, aerospace design and analysis  \cite{mishra2019uncertainty, mishra2017rans, mishra2019estimating, mishra2017uncertainty, thompson2019eigenvector}, application to design under uncertainty (DUU) \cite{demir2023robust, cook2019optimization, mishra2020design, righi2023uncertainties}, virtual certification of aircraft designs \cite{mukhopadhaya2020multi, nigam2021toolset}, etc. 

In spite of its provable successes and widespread adoption the EPM has limitations. An important weakness of the EPM is its reliance on physics only. Based on physics based precepts the EPM can only delineate the \textit{possible} states of turbulence in a specific turbulent flow but not the \textit{probable} states of turbulence for that specific turbulent flow. As an illustration while considering the evolution of the anisotropy of the Reynolds stress tensor the EPM has to weigh all the states including the limiting states equally. This is sub-optimal for a homogeneous flow where the 1- and 2-component limiting states are unlikely. Similarly the EPM applies the same perturbation to all points in the flow domain. The degree of perturbations reflects the degree of discrepancy between the turbulence model predictions and high fidelity data. This discrepancy is not uniform over the entire flow domain and so the degree of perturbations should not be uniform either. These examples highlight the need for a marker function that determines the varying degree (or magnitude) of the perturbations across different regions of a turbulent flow domain and across different turbulent flow instances. 

Researchers have attempted to develop marker functions for the EPM with limited success. We postulate that a completely physics based marker function may not be possible as there is no analytical criterion to predict the discrepancy of a RANS model prediction from the true evolution of the turbulent flow. A Machine Learning (ML) based approach may be able to approximate this marker function to arbitrary accuracy. In the recent past there have been many successful attempts to use ML models and algorithms for problems in fluid flows, turbulence modeling, combustion modeling, etc \cite{duraisamy2019turbulence, brunton2020machine, chung2021data, chung2022interpretable, duraisamy2021perspectives, zhang2015machine}. Many of these attempts focus on the approximation capabilities of the ML model without incorporating the physics knowledge. We attempt to use the basis of the EPM and augment it using domain knowledge informed ML models. Here we identify the absence of non-local modeling information as a key delimiter and utilize an inexpensive Convolutional Neural Network model for the marker function. 

This article is arranged in a sequential manner, where the first section lays out the central question that we are attempting to address the method that we use. In the second section we provide details of the EPM, the correction function and the CNN model utilized in this investigation. The third section gives details of the test cases and data sets used. The fourth section details the methodology used in this study. This is followed by the results of our study and their analysis. The article concludes with a summary of the investigation.

\section{Methods and Details}
\subsection{Eigenspace Perturbation Method (EPM)}
Eddy viscosity models use the concept of a turbulent viscosity to close the evolution equation for the Reynolds stresses. This is also called the Boussinesq Hypothesis \cite{pope2001turbulent}. The instantaneous value of the Reynolds Stresses is assumed to be linearly proportional to the value of the mean rate of strain tensor as

\begin{equation}
    \left\langle u_i u_j\right\rangle=\frac{2}{3} k \delta_{i j}-2 v_{\mathrm{t}}\left\langle S_{i j}\right\rangle,
\end{equation}

where $k$ is the turbulence kinetic energy, $\delta_{i j}$ is the Kronecker delta tensor, $\nu_{t}$ is the eddy viscosity coefficient, and $\left\langle S_{i j}\right\rangle$ is the mean rate of strain. This assumption reduces the complexity of the RANS equations and makes engineering simulations less expensive computationally. But this assumption is very limiting and makes eddy viscosity models inaccurate for complex flows. For example this linear relation asserts that the principal co-ordinates of the mean rate of strain tensor and the Reynolds stress tensor are identical. This is not correct for flows with turbulent separation or re-attachment. This also ignores any effects of the mean rate of rotation on the evolution of the Reynolds stresses, limiting its utility in cases with significant streamline curvature or for rotation-dominated flows.  

To estimate the uncertainties due to eddy viscosity based modeling simplifications the EMP \cite{iaccarino2017eigenspace} introduces perturbations in the spectral representation of the Reynolds stress tensor predicted by the eddy viscosity model as

\begin{equation}\label{Eq:Rij_perturb}
        \left\langle u_{i} u_{j}\right\rangle^{*}=2 k^{*}\left(\frac{1}{3} \delta_{i j}+v_{i n}^{*} \hat{b}_{n l}^{*} v_{j l}^{*}\right).
\end{equation}

Here $\hat{b}_{k l}^{*}$ is the perturbed eigenvalue matrix, $v_{i j}^{*}$ is perturbed eigenvector matrix, $k^{*}$ is the perturbed turbulence kinetic energy. From a modeling perspective, the EPM replaces the linear, isotropic eddy viscosity assumption with the general relationship in which the eddy viscosity is a fourth-order tensor that incorporates the anisotropic nature of turbulent flows\cite{mishra2019theoretical}. 

\subsection{Correction function for RANS predictions}
Computational studies in turbulent flows can use different approaches that differ in the details of the scales that are resolved and the scales that are modeled. In Direct Numerical Simulations (DNS) all the scales of turbulent flow are resolved. DNS is also extremely computationally expensive and not feasible for engineering problems. In Large Eddy Simulations (LES) some scales are resolved and the smaller scales are modeled. While LES is less expensive it is still too computationally expensive for the iterative design methodology used in engineering design. In Reynolds Averaged Navier Stokes (RANS) based modeling all the scales of turbulence are modeled. So RANS based modeling is inexpensive but not as high fidelity as DNS or LES. We use a correction term focusing on the discrepancy between the RANS model predictions and high fidelity DNS data. We refer to this as the correction or the marker function.

In the present study, we aim to utilize an inexpensive CNN-based model to enhance the accuracy of predictions of $k$, crucial in the construction of $k^{*}$ as defined in Equation \ref{Eq:Rij_perturb}. For both the RANS and DNS simulation, we can summarize their results as the function of the perturbed turbulence kinetic energy $k^{*}$: 

\begin{equation}\label{Eq:Marker_Mk_Method}
    k^{*} = f(x,y).
\end{equation}

where $x$ and $y$ are coordinates in a two-dimensional computational domain, and $f$ is the function map from each coordinate $(x, y)$ to $k^{*}$, expressed using a tuple $(x, y, k^{*})$ in simulation results.

Without assuming any form, the correction function for RANS is a mapping between two functions:

\begin{equation}
Z: f^{\mathrm{RANS}}(x,y) \rightarrow f^{\mathrm{DNS}}(x,y)
\end{equation}

with $k^{\mathrm{DNS}} = f^{\mathrm{DNS}}(x, y)$ and $k^{\mathrm{RANS}} = f^{\mathrm{RANS}}(x, y)$, we can rewrite $Z$ as a mapping $\zeta$ between points that comprises $f^{\mathrm{RANS}}$ and $f^{\mathrm{DNS}}$

\begin{equation}
\zeta: (x, y, k^{\mathrm{RANS}}) \rightarrow (x, y, k^{\mathrm{DNS}})
\end{equation}

Considering the model error for RANS and DNS in terms of kinetic energy, we have

\begin{equation}
   p^{\text {RANS }}\left(K_g \mid x, y\right)=p\left(k_g=k^{\text {RANS }} \mid x, y\right)
\end{equation}

\begin{equation}
    p^{\text {DNS }}\left(K_g \mid x, y\right)=p\left(k_g=k^{\text {DNS }} \mid x, y\right)
\end{equation}

where $K_g$ is the unknown ground truth of kinetic energy at $(x, y)$.

Kinetic energy resulted from DNS simulation results $p^{\mathrm{RANS}}$ can be estimated with kinetic energy from RANS simulation $p^{\mathrm{DNS}}$ and its correction function $g$ as

\begin{equation}
p^{\mathrm{DNS}}\left(K_g \mid x, y\right)=g\left(k^{\mathrm{RANS}}, x, y\right) p\left(k^{\mathrm{RANS}} \mid x, y\right)
\end{equation}

As $k^{\mathrm{DNS}} = f^{\mathrm{DNS}}(x, y)$ and $k^{\mathrm{RANS}} = f^{\mathrm{RANS}}(x, y)$, at each $x$, we have that $k_x^{\mathrm{DNS}} = f_x^{\mathrm{DNS}}(y)$ and $k_x^{\mathrm{RANS}} = f_x^{\mathrm{RANS}}(y)$, assuming both $f_x^{\mathrm{RANS}}$ and $f_x^{\mathrm{DNS}}$ are continuous. In other words, we can learn $\hat{g}$ with paired $(\mathbf{k}_{x,y,\delta}^{\mathrm{RANS}}, \mathbf{k}_{x,y,\delta}^{\mathrm{DNS}})$.

\subsection{CNN-based Correction Function}
This investigation employs a one-dimensional convolutional neural network (1D-CNN) to learn the correction function $\hat{g}$ from paired RANS and DNS simulation estimated kinetic energy $(\mathbf{k}_{x,y,\delta}^{\mathrm{RANS}}, \mathbf{k}_{x,y,\delta}^{\mathrm{DNS}})$. As our approximated correction function $\hat{g}$ depends on the neighbor of $k^{\mathrm{RANS}}$ and coordinates $(x, y)$ are only used to group neighbors of $k^{\mathrm{RANS}}$, we grouped simulation data by $x$ and transformed $(y, k)$ at $x$ into $\mathbf{k}_{x,y,\delta}^{\mathrm{RANS}}$ via a rolling window parameterized by window size. Our 1D-CNN has four-layers and in total 86 parameters: a single model for all zones at any $x$ to correct RANS towards DNS.

\section{Turbulent Flow Test Cases Details}
This investigation uses two datasets for training the CNN model: 
\begin{enumerate}
    \item an in-house paired RANS and DNS dataset
    \item an open source dataset made of paired RANS and DNS data.
\end{enumerate}

\begin{figure} \label{Fig:Cgrid}
\centerline{\includegraphics[width=6in]{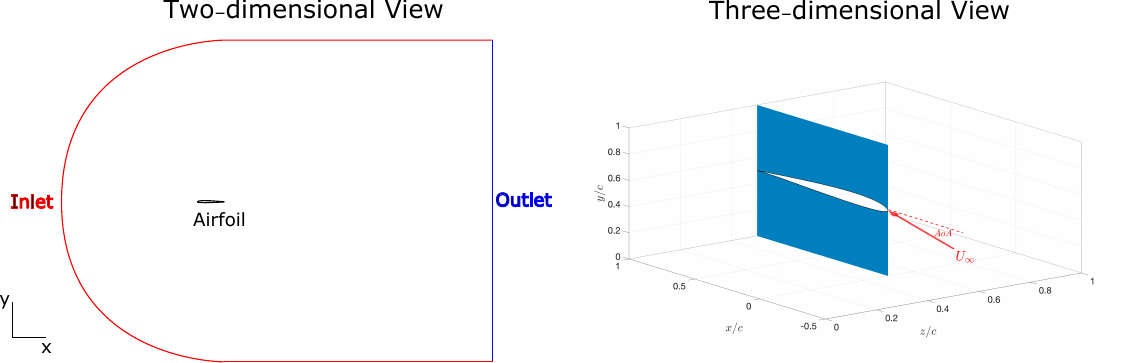}}
\caption{The computational domain and boundary conditions for the SD7003 airfoil case: far field, outlet, and no-slip walls. A three-dimensional version of the computational domain is provided with freestream ($U_{\infty}$) encountering the leading edge at $8^{\circ}$ AoA.}
\end{figure}

For the first dataset, the DNS data was generated by simulating flow over an SD7003 airfoil at $8$ degrees angle of attack, Reynolds number based on the cord length of $Re_{c} = 60000$ and a Mach number (Ma) of $0.2$. In this setting, for the spanwise direction, a periodic boundary condition is adopted, and the spanwise domain extends to a length of $0.25$ times the chord length. The DNS mesh has $1200$ nodes in the spanwise direction, $800$ nodes in the axial direction, and $96$ nodes in the last direction. The detailed information about the DNS dataset and its generation is available in the thesis of \cite{zhang2021turbulent}.

For the RANS data corresponding to the first paired dataset, RANS simulation is carried out for airflow around the SD7003 airfoil. Here the freestream inlet velocity ($U_{\infty}$) transitioned to turbulence on the suction side of the airfoil at an angle of attack (AoA) of $8$ degrees relative to the freestream, corresponding to $Re_{c} = 60,000$. The solution domain has a two-dimensional C topology grid with a resolution of 389 (along the streamwise axis) by 280 (along the wall normal axis) by 1 (along the spanwise axis) control volumes shown in Fig. \ref{Fig:Cgrid}. The grid approximates that of a prior referenced case (768 by 176) in \cite{catalano2011rans}. To ascertain the effect of grid resolution on the RANS solution a grid independence study focusing on the near wall zone was carried out. The study exhibits that employing a finer mesh did not substantially change the accuracy of the RANS predictions. The influence on the mean velocity profile as well as Reynolds shear stress profile was less than $1\%$. In the current manuscript the plots for grid convergence are excluded for brevity. For this investigation, we utilize the less resolved mesh ($389 \times 280$) for simulations. A low inlet freestream turbulence intensity level is assumed and set at $Tu = 0.03\%$. The outlet conditions use a zero gradient boundary condition for variables including mean streamwise velocity in the x-direction, mean velocity in the y-direction, turbulence kinetic energy, and pressure. As is customary no slip boundary conditions were used at the walls. The initial grid nodes in the wall-normal direction were positioned at $y^{+} \approx 1$ within the turbulent boundary layer utilizing over $20$ control volumes. 

The RANS based transition model of Langtry and Menter \cite{langtry2009correlation} is used for the closure of the continuity and the momentum equations (detailed in \ref{appendixA}. The transport equations are discretized using finite volumes and a staggered mesh. Spatial discretization used second-order upwind scheme and gradients are estimated with the Gaussian linear scheme. With OpenFOAM the unsteady PIMPLE solver is used for coupling pressure and velocity fields. We use a maximum Courant number of $0.6$. Regardless OpenFOAM dynamically adjusts the time step to maintain the specified maximum Courant number. 

Convergence tracking was carried out by monitoring both residuals along with the lift and drag coefficients with time ($T$). On convergence, specifically at $T \approx 0.3$ or normalized time $T = TU_{\infty}/c \approx 6.75$, where $U_{\infty}$ is the freestream velocity magnitude, the energy and momentum residuals had decreased by over four orders of magnitude. At the same time the lift and drag coefficients exhibited negligible variation. Similar behavior was seen in the study by Catalano and Tognaccini \cite{catalano2011rans} for a low Reynolds number flow over a SD7003 airfoil at AoA = $10^\circ$. Our sampling started at $T = 0.6$ (twice the time of convergence) and finished at $T=1.4$, involving roughly $35000$ iterations over the simulations.

\subsubsection{RANS/DNS approach for 2D channel flow over periodic hills}
\begin{figure} 
\centerline{\includegraphics[width=6in]{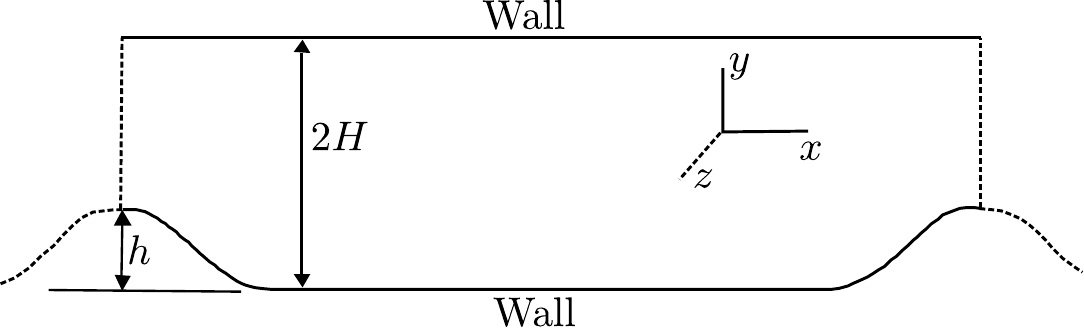}}
\caption{The 2D periodic hills computational domain used in this investigation.}
\label{fig:periodichill.pdf}
\end{figure}

This section introduces the outline of the approach for obtaining the \texttt{Voet} dataset. Detailed information is given in the prior publication of \cite{voet2021hybrid}. The Reynolds number is set at $5600$ for the RANS simulations and the DNS simulation. At this regime of Reynolds numbers, the spectrum of turbulent scales is broad enough to make a RANS approach possible while the DNS simulation is feasible computationally. The DNS simulations are carried out on a grid sized $L_{x} \times L_{y} \times \times L_{z} = 9.000h \times 3.036h \times 4.500h$, where $h$ represents the hill height, as shown in Figure \ref{fig:periodichill.pdf}. The coordinate system is orthogonal and the $x$-axis is aligned with the stream-wise direction, the $y$-axis with the vertical direction, and the $z$-axis with the spanwise direction \cite{voet2021hybrid}. A structured mesh of $512 \times 257 \times 128$ mesh nodes in the directions is used. A $y$-stretched mesh is used, with refinement around the walls. The spatial resolution exceeds that of prior numerical studies conducted at the same Reynolds number. The boundary conditions used are a periodic interface in both the streamwise and spanwise directions, and a non-slip condition at the top and bottom walls of the channel. The initial setup of the horizontal velocity profile in the channel uses a Poiseuille flow condition given by

\begin{equation} \label{Eq:DNS_initial_profile}
    u(y)=U_0\left(1-\left(\frac{y}{H}\right)^2\right),
\end{equation}

where $H$ is the semi channel height and $U_{0}$ the center line velocity. The DNS is conducted using the high-order flow solver, Incompact3d \cite{laizet2009high,laizet2011incompact3d}. More information about the DNS dataset for flow over periodic hills is given in \cite{voet2021hybrid,xiao2020flows}.

Two dimensional RANS simulations are carried out using the Star-CCM+ software. The computational domain, as shown in Figure \ref{fig:periodichill.pdf}, is represented as a two-dimensional domain while maintaining the size of the domain at $L_{x} \times L_{y} = 9h \times 3.036h$. For the upper and lower walls, we use a no slip boundary condition. For the inlet and outlet of the fluid domain, we use a periodic boundary condition. The $k-\omega$ Shear Stress Transport (SST) turbulence model is used \cite{menter2003ten}. The RANS simulations run for over $10,000$ iterations, with residuals below $4 \times 10^{-10}$ for the continuity, U and V momentum, turbulent kinetic energy, and turbulent dissipation rate equations, as shown in the reference test case ($\alpha = 1$, $\gamma = 1$). For a given Reynolds number, the flow over periodic hills is centrally influenced by two factors: the steepness and the spacing of the hills. The two parameters, $\alpha$ and $\gamma$ as defined, are used to manage the uncertainty of the geometry. The parameter $\alpha$ changes the steepness of the hills, so altering the adverse pressure gradient on the flow when moving down the hill. It is expected to impact the separation point. The parameter $\gamma$ changes the spacing between successive hills, and is expected to influence the reattachment point. By varying the values of $\gamma$ and $\alpha$, $7$ data points are obtained from DNS studies and $30$ data points are obtained using RANS simulations. The details on this dataset are given in \ref{appendixB}. 

\section{Numerical Experiments}
The primary objective of this study is to use computationally feasible convolutional neural networks to modulate the application of the Eigenspace Perturbation Framework. Our specific focus is to develop a marker function that can control the magnitude of the eigenvalue perturbation while varying it over the computational domain. With this objective we use the CNN-based approach to approximate this correction function for RANS simulations on two distinct datasets: an in-house RANS/DNS \cite{zhang2021turbulent,chu2022model} dataset for an SD7003 airfoil at $8^\circ$ angle of attack (referred to as \texttt{SD7003} in the following text) and the public RANS/DNS dataset \cite{voet2021hybrid} for two-dimensional channel flow over periodic hills.

\subsection{Data Flow and Model Configuration} \label{sec:VoetData}
\begin{figure}[h!]
    \centering
    \includegraphics[width=\linewidth]{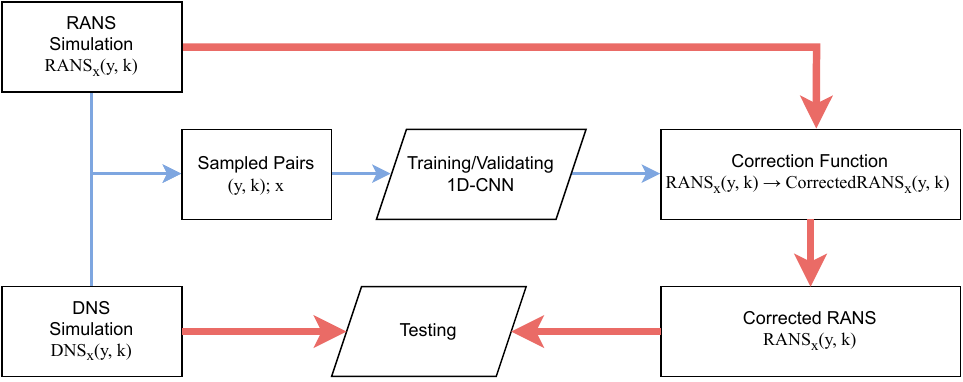}
    \caption{The data flow and methodology used in this investigation. The blue path is the training path,the red path is the validation path.} 
    \label{fig:data-flow.pdf}
\end{figure}

For the two set cases outlined earlier, we set up experiments of the 1D-CNN model following the simulation data generation as outlined in Figure \ref{fig:data-flow.pdf}. The DNS data is treated as the ground truth to evaluate the 1D-CNN corrected RANS prediction. We use the non-corrected RANS prediction as a control sample. As the DNS and RANS often use different grids, we select $x$-coordinates common to both DNS and RANS meshes. We split $x$-coordinate grouped pairs of $(\mathbf{k}_{x,y,\delta}^{\mathrm{RANS}}, \mathbf{k}_{x,y,\delta}^{\mathrm{DNS}})$ into a training set and a validation set by the  key $x$. We utilize a 80\%--20\% split of the raw dataset for training and testing purposes. 

To train the model, we use the Mean Absolute Error (MAE) as the objective function. In contrast to the Mean Squared Error (MSE or $L_2$ loss) the MAE (or the $L_1$ loss) does not penalize incorrect predictions as heavily and better final models were developed in this study using the MAE loss as the objective function. The MAE loss is computed for the uncorrected RANS and the MAE loss of 1D-CNN corrected RANS. These are compared to exhibit the efficacy of our approach.

\begin{figure*}[h!]
    \centering
    \includegraphics[width=\linewidth]{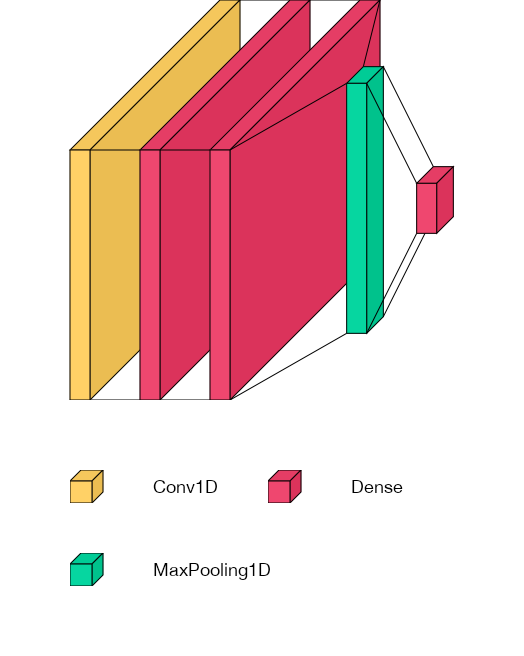}
    \caption{Architecture of the Convolutional Neural Network used in this study} 
    \label{fig:model-layered-view.pdf}
\end{figure*}

Figure \ref{fig:model-layered-view.pdf} reports the architecture of the convolutional neural network used in this study. We arrange the input from RANS with a window size of 11 for each DNS reading (window size of 1). The 1-dimensional convolutional layer is configured with a kernel size of 3, and a stride of 1. This layer is able to learn representations of the high-level features from the input data through convolution operations. After the convolutional layer we have two dense layers interpreting the feature extracted by the prior layers. Following these, we have a Pooling layer (max pool). The model training requires the use of the Adam optimizer with a learning rate of $0.001$ for 800 epochs with a batch size of 10.

\section{Results, Analysis and Inference}

As outlined in the data flow diagram of the prior section, we apply and verify the performance of the CNN approach at all paired (RANS, DNS) datasets at the common $x$ locations. In this section, we report the results are four representative $x$ axis spans. These locations are spanning across the separated region: for the \texttt{SD7003} dataset with RANS and DNS based on the airfoil geometry, $x/c = 0.17, 0.25, 0.32, 0.44$; and for the \texttt{Voet} dataset based on the two-dimensional periodic hills, $x/H = 0, 0.035, 1.961, 4.885, 6.847$.

\begin{figure}
    \centering
    \includegraphics[width=\linewidth]{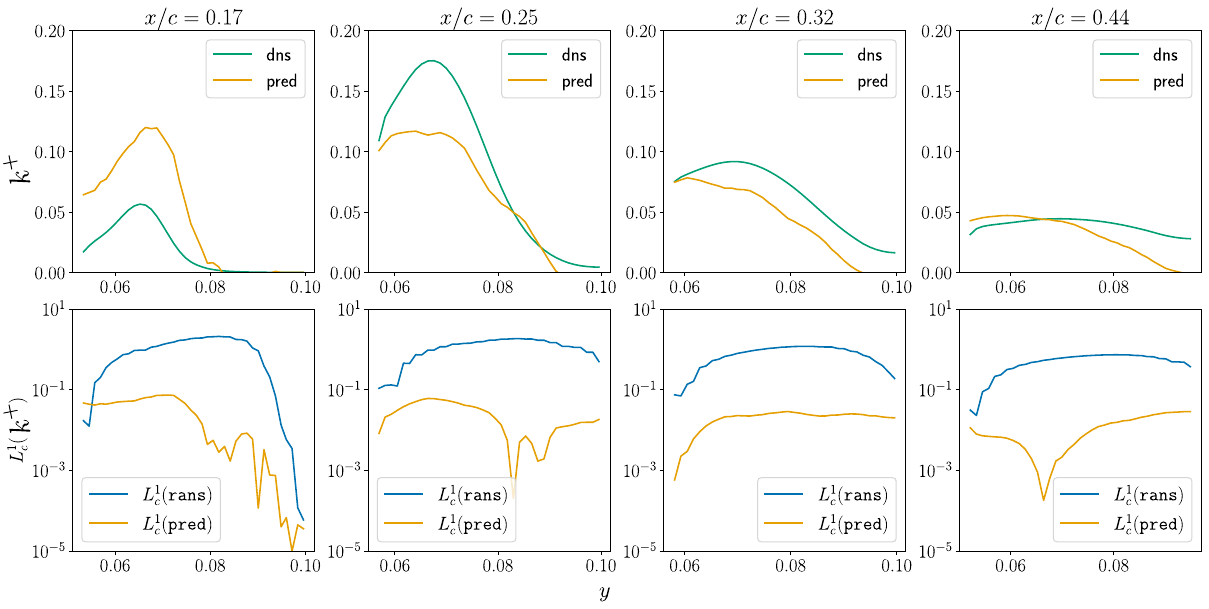}
    \caption{Results for \texttt{SD7003} dataset. DNS-based CNN prediction for normalized turbulence kinetic energy. First row: CNN predicted DNS (\texttt{pred}) compared with ground truth (\texttt{dns}). Second row: Validation of 1D-CNN by comparing L1 loss between $L^1_c(\texttt{rans})$ and $L^1_c(\texttt{pred})$.}
    \label{fig:rans-predicted-dns-main4-viz-loss.pdf}
\end{figure}

\begin{figure} 
    \centering
    \includegraphics[width=\linewidth]{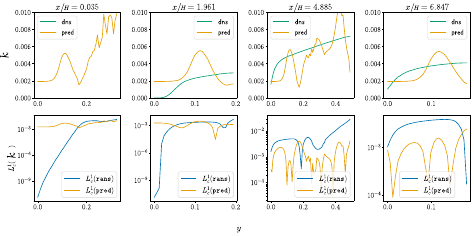}
    \caption{Results for the \texttt{Voet} dataset for periodic hills. DNS case2-based CNN prediction for dimensional turbulence kinetic energy. First row: corrected DNS case 2 (pred) compared with ground truth (dns) case 7. For DNS case 2: $\alpha = 0.8$ and $\gamma = 1.0$; DNS case 7: $\alpha = 1.2$ and $\gamma = 1.0$. Second row: Validation of 1D-CNN by comparing L1 loss between $L^1_c(\texttt{rans})$ and $L^1_c(\texttt{pred})$. }
    \label{fig:case2-based-dns2}
\end{figure}

In the series of figures \ref{fig:rans-predicted-dns-main4-viz-loss.pdf} - \ref{fig:case2-based-dns2} we report the results for test case of a flow over SD7003 airfoil and 2D periodic hills. From Figures \ref{fig:rans-predicted-dns-main4-viz-loss.pdf} - \ref{fig:case2-based-dns2}, the CNN predicted profiles in the first row are smoothed with the moving average with a window size of six steps. The CNN prediction for the turbulent kinetic energy profile closely matches the ground truth DNS in spite of the small size of the training data. The turbulent kinetic energy is unscaled for Figure \ref{fig:case2-based-dns2}, and normalized turbulence kinetic energy is used for \ref{fig:rans-predicted-dns-main4-viz-loss.pdf}, i.e., $k^{+} = k/U_{\infty}^2$. In Fig. \ref{fig:rans-predicted-dns-main4-viz-loss.pdf}, the CNN predicted DNS profiles approximate the qualitative features of ground truth DNS profiles. A discrepancy is observed at $x/c = 0.7$ within the separated region. As the flow travels further downstream, the CNN-predicted DNS profiles are closer to the ground truth, showing improved accuracy of the CNN correction function in the region. The corresponding $L^1_c(\texttt{pred})$ value shows a substantial decrease for all positions, approximately by two orders of magnitude, indicating enhanced accuracy in the CNN predictions. \ref{appendixC} shows an example of the application of the CNN correction function on the suction side of the SD7003 airfoil to predict turbulent kinetic energy. 

For this investigation, we carried out additional analysis on the public RANS/DNS dataset \cite{voet2021hybrid} to train the CNN model. In Fig. \ref{fig:case2-based-dns2}, we use this model, trained on the case two from the RANS/DNS dataset \cite{voet2021hybrid}, to predict the $k$ profiles of DNS for case seven from the RANS/DNS dataset. For case 2, $\alpha = 0.8$ and $\gamma = 1.0$, while for case 7, $\alpha = 1.2$ and $\gamma = 1.0$, where $\alpha$ changes the hill steepness while $\gamma$ changes the successive spacing of the hills (the reader is directed to \ref{appendixB} for details). Turbulent kinetic energy is not scaled for this comparison. From Fig. \ref{fig:case2-based-dns2}, the CNN predicted DNS profiles approximate well the ground truth DNS profiles at all positions, except for $x/H = 0.034$, where a discrepancy is observed. This is due to the highly turbulent state of the flow at this location, caused in part by the high degree of separation. The associated $L^1_c(\texttt{pred})$ shows large discrepancy close to the wall at $x/H = 0.035$ and $x/H = 1.961$. This highlights the necessity for our CNN-based correction function to concentrate on enhancing predictive accuracy, particularly within a strongly separated region. As the flow proceeds downstream, the overall $L^1_c(\texttt{pred})$ value remains approximately 1-2 orders of magnitude lower than $L^1_c(\texttt{rans})$, suggesting improved predictions. A central limitation of ML models is their inability to generalize from their training dataset to other, different flows. Our experiments exhibit that this CNN model is robust and retains good predictions across different training data sets.

In this investigation we use a CNN model to approximate the correction function that reduces the discrepancy between RANS simulation and DNS data. This correction function reflects the discrepancy in the RANS simulation results and so it  acts as a spatially varying marker function for the eigenvalue perturbations by providing guidance on the magnitude of perturbation for $k^{*}$ required in Eqn. \ref{Eq:Rij_perturb}. We experiment and apply this method on two diverse datasets.

While the datasets corresponding to very different flows, both include flow separation and re-attachment, leading to the presence of separation bubbles. In both the flow test cases, the CNN model's correction function can significantly reduce the discrepancy between RANS predictions and DNS-simulations.

In both the test cases, our model was trained on a relatively small dataset. Additionally the CNN model was robust to covariate shifts where the model trained on one flow test case gave consistently good predictions on the other test flow case. At present there are few studies for using ML models to develop marker functions that indicate the degree of discrepancy in RANS predictions, thus assisting in the Eigenspace Perturbation Method. Recently, the study of Chu \textit{et al.} \cite{chu2022model} analyzed the use of polynomial regression to the perturbed turbulence kinetic energy. This CNN model correction method has implications on practical applications such as, to be coupled to the EPM. The EPM has been implemented within the OpenFOAM framework to construct a marker function for the perturbed turbulence kinetic energy \cite{chu2022model}. In future work, this CNN model correction method will be utilized as a spatially varying marker function to guide the degree of eigenvalue perturbation.

\section{Summary, Conclusions and Future Work}
This investigation seeks to use deep convolutional neural networks to formulate a spatially varying marker function that can modulate the magnitude of the eigenvalue perturbation in the EIgenspace Perturbation Framework. In the recent past there have been investigations that seek to utilize Ml models to modulate and improve the Eigenspace Perturbation Framework \cite{heyse2021data, heyse2021estimating, matha2023evaluation}. However the present study is the first to analyze the projection from RANS prediction space to the space of DNS data using the CNN models. Our use of convolutions allows the possibility of the inclusion of non-local information in the uncertainty estimation. The results show that the CNN models learn the discrepancy between RANS simulations and DNS data, resulting in a surrogate model for the marker function.

\appendix

\section{The correlation-based transition model of Langtry and Menter}
\label{appendixA}

If the flow can be assumed to be 2-dimensional, unsteady and incompressible, the governing equations can be simplified to
\begin{equation} \label{p_Continuity}
   \frac{\partial \left\langle U_{i} \right\rangle}{\partial x_{i}}=0,
\end{equation}
\begin{equation} \label{p_Momentum}
   \frac{ D \left\langle U_{j}\right\rangle}{\mathrm{Dt}}=-\frac{1}{\rho} \frac{\partial \left\langle P \right\rangle}{\partial x_{j}}+\nu \frac{\partial^{2} {\left\langle U_{j} \right\rangle}}{\partial x_{i} \partial x_{i}}-\frac{\partial \left\langle u_{i} u_{j}\right\rangle}{\partial x_{i}}
\end{equation}

where $\left\langle \ \right\rangle$ represents time-averaging operation. $\rho$ is density, $\left\langle P \right\rangle$ is the time-averaged pressure, and $\nu$ is the kinematic viscosity. $\left\langle U_{i}\right\rangle$ are the time-averaged velocity components. Note that the Reynolds stress term in Eqn. \ref{p_Momentum} is unknown and is approximated using a turbulence model. We use a two-equation linear eddy viscosity model of  Langtry and Menter \cite{menter1993zonal,hellsten1998some,menter2001elements,menter2003ten} that is used in engineering, for example aerospace industry. This RANS model is based on the Boussinesq turbulent viscosity hypothesis as follows:

\begin{equation}\label{Eq:noMark_uiuj}
    \left\langle{u_{i} u_{j}}\right\rangle=\frac{2}{3} k \delta_{i j}-2 \nu_{\mathrm{t}} \left\langle S_{i j} \right\rangle,
\end{equation}

where $k$ is the turbulence kinetic energy, $\delta_{i j}$ is the Kronecker delta, $\nu_\mathrm{t}$ is the turbulent viscosity, and $\left\langle S_{i j} \right\rangle$ is the rate of mean strain tensor. In Eq. \ref{Eq:noMark_uiuj}, the deviatoric component is 

\begin{equation}\label{Eqn:Bou_Ani_Tensor}
\begin{aligned}
a_{i j} & \equiv\left\langle u_{i} u_{j}\right\rangle-\frac{2}{3} k \delta_{i j} \\
&=-\nu_{\mathrm{t}}\left(\frac{\partial\left\langle U_{i}\right\rangle}{\partial x_{j}}+\frac{\partial\left\langle U_{j}\right\rangle}{\partial x_{i}}\right) \\
&=-2 \nu_{\mathrm{t}} \left\langle S_{i j} \right\rangle.
\end{aligned}
\end{equation}

The transition model by Langtry and Menter \cite{menter1993zonal,hellsten1998some,menter2001elements,menter2003ten} does not approximate the physics of transition. instead, the transition process is represented via correlations. There are two additional transport equations introduced in this transition model for the intermittency and the transition production. The first transport equation for the intermittency $\lambda$ is:

\begin{equation}\label{Eq:gamma}
    \frac{\partial( \gamma)}{\partial t}+\frac{\partial\left( \left\langle U_{j} \right\rangle \gamma\right)}{\partial x_{j}}=P_{\gamma}-E_{\gamma}+\frac{\partial}{\partial x_{j}}\left[\left(\nu+\frac{\nu_{t}}{\sigma_{f}}\right) \frac{\partial \gamma}{\partial x_{j}}\right].
\end{equation}

In Eq. \ref{Eq:gamma}, intermittency lies between zero to one. In the freestream, the intermittency is assumed to be one to improve robustness across applications. The transition production term is:

\begin{equation}\label{Eq:gamma_source}
    P_{\gamma 1}=F_{\text {length }} c_{a 1} S\left[\gamma F_{\text {onset }}\right]^{0.5}\left(1-c_{e 1} \gamma\right),
\end{equation}

where $S$ is the strain-rate magnitude. Both $F_{\text {length }}$ and $F_{\text {onset }}$ are dimensionless functions to modulate the intermittency in the boundary layer. $F_{\text {length }}$ is a correlation to modulate the length of the transition region, and $F_{\text {onset}}$ engenders the onset of transition through the local vorticity Reynolds number \cite{menter2002transition} $Re_{\nu}$.  So the transition onset is formulated as:

\begin{equation}\label{Eq:Re_nu}
    R e_{\nu}=\frac{ y^{2} S}{\nu},
\end{equation}

\begin{equation}\label{Eq:Re_thetac}
    F_{\text {onset1 }}=\frac{R e_{v}}{2.193 \cdot R e_{\theta c}},
\end{equation}

\begin{equation}
    F_{\text {onset } 2}=\min \left(\max \left(F_{\text {onset1 } 1}, F_{\text {onset1 }}^{4}\right), 2.0\right),
\end{equation}

\begin{equation}
    R_{T}=\frac{ k}{\nu \omega},
\end{equation}

\begin{equation}
    F_{\mathrm{onset} 3}=\max \left(1-\left(\frac{R_{T}}{2.5}\right)^{3}, 0\right),
\end{equation}

\begin{equation}
    F_{\textit {onset}}=\max \left(F_{\text {onset} 2}-F_{\text {onset3}}, 0\right).
\end{equation}

The final expression in Eq. \ref{Eq:gamma_source} modulates the  maximum the intermittency, i.e., the value of intermittency is always lesser than one. In Eqn. \ref{Eq:Re_nu}, the constant $c_{e 1}$ of $1.0$ is used, and $y$ is the wall distance. In Eqn. \ref{Eq:Re_thetac}, $R e_{\theta c}$ is the critical Reynolds number at which turbulence grows in the boundary layer. $R e_{\theta c}$ precedes the transition Reynolds number $\tilde{\operatorname{Re}}_{\theta t}$ where the velocity profile deviates from the laminar profile. Like $F_{length}$, $R e_{\theta c}$ is an empirical correlation. Both the $F_{length}$ and $R e_{\theta c}$ \cite{langtry2009correlation} correlations are functions of $\tilde{\operatorname{Re}}_{\theta t}$ \cite{langtry2009correlation}. 

In Eq. \ref{Eq:gamma}, the transition destruction source term is:

\begin{equation}
    E_{\gamma}=c_{a 2} \Omega \gamma F_{\text {turb }}\left(c_{e 2} \gamma-1\right),
\end{equation}

where $\Omega$ represents the vorticity magnitude. The destruction source term enables reduction of intermittency to zero in the laminar boundary layer. Once the transition criteria are no longer met in the $F_{onset}$ function, the model anticipates relaminarization by setting the intermittency value to zero.  The coefficients $C_{a 2}$ of $0.06$ represents the intensity of the destruction term, ensuring it remains smaller than the source term. The constant $C_{e 2}$ of $50$ modulates the minimum threshold of intermittency. $F_{turb}$ \cite{langtry2009correlation} is used to deactivate the destruction source term outside the laminar boundary layer or inside the viscous sublayer. 

This investigation focuses on transition flow separation over a SD7003 airfoil. A benefit of Langtry and Menter's correlation-based transition model \cite{langtry2009correlation}  \cite{menter2004correlation,menter2006correlation,langtry2009correlation} is its capability to predict the separation-induced transition. The modification to the intermittency for separation-induced transition is: 

\begin{equation}\label{Eq:gamma_sep}
    \gamma_{s e p}=\min \left(s_{1} \max \left[0,\left(\frac{\operatorname{Re}_{\nu}}{3.235 \operatorname{Re}_{\theta c}}\right)-1\right] F_{\text {reatach}}, 2\right) F_{\theta t},
\end{equation}

\begin{equation}
     F_{\text {reatach}}=e^{-\left(\frac{R_{T}}{20}\right)^{4}},
\end{equation}

\begin{equation}\label{Eq:gamma_eff}
     \gamma_{e f f}=\max \left(\gamma, \gamma_{s e p}\right),
\end{equation}

\begin{equation}
     s_{1}=2.
\end{equation}

The constant $s_{1} = 2$ modulates the size of the separation bubble. $F_{reattach}$ deactivates the separation based transition when the viscosity ratio is high enough for reattachment to occur. The effective intermittency value $\gamma_{eff}$ is determined from Eq. \ref{Eq:gamma}, except in the separation-induced transitional boundary layer. This leads to an excessive generation of turbulent kinetic energy, forcing the boundary layer reattachment. If the value of $s_{1}$ is increased, the length of the separated region decreases, and vice versa.

The second transport equation for the transition Reynolds number is:

\begin{equation}\label{Eq:rethetatloc}
    \frac{\partial\left( \tilde{\operatorname{Re}}_{\theta t}\right)}{\partial t}+\frac{\partial\left( \left\langle U_{j} \right\rangle \tilde{\mathrm{Re}_{\theta t}}\right)}{\partial x_{j}}=P_{\theta t}+\frac{\partial}{\partial x_{j}}\left[\sigma_{\theta t}\left(\nu+\nu_{t}\right) \frac{\partial \tilde{\mathrm{Re}}_{\theta t}}{\partial x_{j}}\right],
\end{equation}
 
where $P_{\theta t}$ represents the source term. Outside of the boundary layer, $P_{\theta t}$ causes the transported scalar $\tilde{\operatorname{Re}}_{\theta t}$ to match the local value of $\operatorname{Re}_{\theta t}$, calculated via a correlation \cite{langtry2009correlation}. The source term is formulated as
 
 \begin{equation}\label{Eq:rethetatP}
    P_{\theta t}= \frac{c_{\theta t}}{t}\left(\operatorname{Re}_{\theta t}-\tilde{\operatorname{Re}_{\theta t}}\right)\left(1.0-F_{\theta t}\right),
\end{equation}
 
 \begin{equation}
     t = \frac{500\nu}{U^2},
 \end{equation}
 
where $t$ is a time scale derived from dimensional analysis, and $U = \sqrt{U_{1}^2 + U_{2}^2 + U_{3}^2}$ the mean velocity. The blending function $F_{\theta t}$ deactivates the source term within the boundary layer. The blending function $F_{\theta t}$ is
 
 \begin{equation}\label{Eq:Fthetat}
    F_{\theta t}=\min \left(\max \left(F_{\text {wake}} \cdot e^{-\left(\frac{y}{\delta}\right)^{4}}, 1.0-\left(\frac{\gamma-1 / c_{e 2}}{1.0-1 / c_{e 2}}\right)^{2}\right), 1.0\right),
\end{equation}

where 

\begin{equation}\label{Fthetat_theta_delta}
    \theta_{B L}=\frac{\tilde{\operatorname{Re}_{\theta t}} \nu}{ U}; \quad \delta_{B L}=\frac{15}{2} \theta_{B L}; \quad \delta=\frac{50 \Omega y}{U} \cdot \delta_{B L};
\end{equation}

\begin{equation}\label{Eq:Fthetat_reomega_wake}
    \operatorname{Re}_{\omega}=\frac{ \omega y^{2}}{\nu}; \quad F_{\text {wake}}=e^{-\left(\frac{\mathrm{Re}_{\omega}}{1 \times 10^{5}}\right)^{2}}.
\end{equation}

The $F_{wake}$ function suppresses this blending function in the wake regions. The model constants $c_{\theta t} = 0.03$ and $\sigma_{\theta t} = 2.0$ are used to modulate the source term and the diffusion coefficient.

The transition model \cite{menter2004correlation,menter2006correlation,langtry2009correlation} interacts with the SST $k - \omega$ \cite{menter1993zonal,hellsten1998some,menter2001elements,menter2003ten} turbulence model as:

\begin{equation}\label{Eq:kEqn}
    \frac{\partial}{\partial t}( k)+\frac{\partial}{\partial x_{j}}\left( \left\langle U_{j} \right\rangle k\right)=\tilde{P}_{k}-\tilde{D}_{k}+\frac{\partial}{\partial x_{j}}\left(\left(\nu+\sigma_{k} \nu_{t}\right) \frac{\partial k}{\partial x_{j}}\right),
\end{equation}

\begin{equation}
    \tilde{P_{k}} = \gamma_{eff}P_{k}; \quad \tilde{D_{k}} = \min\left(\max\left(\gamma_{eff},0.1\right),1.0\right)D_{k};
\end{equation}

\begin{equation}
    R_{y} = \frac{y\sqrt{k}}{\nu}; \quad F_{3} = e^{\left(-\frac{R_{y}}{120}\right)^{8}}; \quad F_{1} = \max\left(F_{lorig},F_{3}\right).
\end{equation}  

where $P_{k}$ and $D_{k}$ are the production and destruction terms in the SST $k-\omega$ turbulence model \cite{menter1993zonal,hellsten1998some,menter2001elements,menter2003ten}. The effective intermittency term in Eq. \ref{Eq:gamma_eff} modulates the source term in the turbulence kinetic energy transport equation. The bending function $F_{lorig}$ for the SST $k-\omega$ model \cite{menter1993zonal,hellsten1998some,menter2001elements,menter2003ten} is developed exclusively for the turbulent boundary layers and ceases operation at the center of the laminar boundary layer. To capture transition, the model needs to be active in laminar and in transitional boundary layers. Therefore, the transition model proposed by Langtry and Menter \cite{menter2004correlation,menter2006correlation,langtry2009correlation} defines the blending function $F_{1}$ to enable its activation within a laminar boundary layer.

\section{The Voet Data Set}
\label{appendixB}
The open source data of Voet \textit{et al.} \cite{voet2021hybrid} was used to train the Convolutional Neural Network model in this study. In this study, we use the data 
stored in the \texttt{rms\_files1.dat} files. The featture or column names from left to right in \texttt{rms\_files1.dat} file are \texttt{x}, \texttt{y}, \texttt{uumean}, \texttt{vvmean}, \texttt{wwmean}, and \texttt{p}, respectively. Physically they represent $x$ coordinate, $y$ coordinate, time-averaged turbulent velocity component in $x$, $y$, and $z$ directions. The DNS data of \texttt{x\_dns, y\_dns, k\_dns} are used in training our CNN model, i.e., \texttt{x\_dns = x}, \texttt{y\_dns = y}, and \texttt{k\_dns = 1/2 (uumean, vvmean, wwmean)}.

The RANS data are shared with the names of \texttt{Results\_bump(\textbackslash d+).csv}, where \texttt{(\textbackslash d+)} is the case index written in the syntax of regular-expression. The column names are clearly given: \texttt{"Velocity[i] (m/s)"} as \texttt{x}, \texttt{"Velocity[j] (m/s)"} as \texttt{y}, and \texttt{"turbulence kinetic Energy (J/kg)"} as \texttt{k}. 

According to Voet \textit{et al.} \cite{voet2021hybrid}, both the  DNS and the  RANS simulations should have their $x$ and $y$ normalized by the characteristic length hill height $H = 0.028m$. Therefore, the data format of \texttt{x\_{$RANS \textbackslash DNS$}} and \texttt{y\_{$RANS \textbackslash DNS$}} is used in training the CNN model. Voet \textit{et al.} \cite{voet2021hybrid} defined two main parameters $\alpha$ and $\gamma$ to influence the flow over the periodic hills. It is noted that the DNS and RANS cases are not paired by case index; instead corresponding DNS and RANS cases can only be determined using $\alpha$ and $\gamma$. We compared these two parameters across DNS cases and RANS cases, and found that DNS Case1 is matched with RANS Case18, similarly, DNS Case2 with RANS Case6, DNS Case3 with RANS Case16, DNS Case4 with RANS Case13, DNS Case5 with RANS Case12, DNS Case6 with RANS Case24, DNS Case7 with RANS Case30.

\section{Application of the lightweight CNN-based correction function on UQ for an SD 7003 airfoil}\label{appendixC}
Once trained, the Convolutional Neural Network model from this study can be applied to predict the correction function for different flow cases. Here we outline its application for a specific flow case.

\begin{figure*}[t]
         \centering
         \includegraphics[width=\textwidth, trim={2.4mm 0 4mm 0},clip]{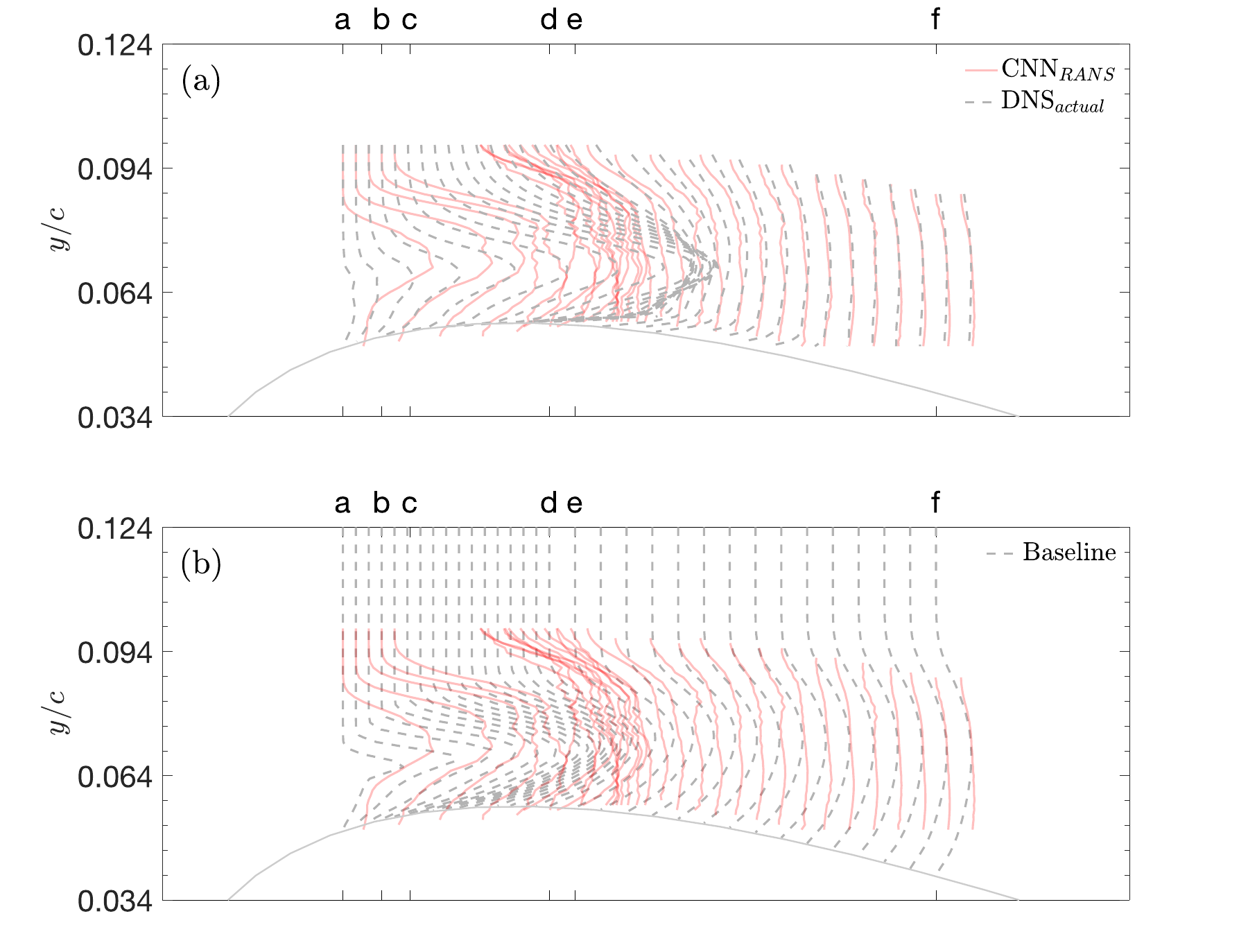}
        \caption{(a) CNN corrected RANS (\texttt{CNN\_{RANS}}) {(solid-dotted lines)} of the normalized perturbed turbulence kinetic energy and (b) ground truth (\texttt{DNS\_{actual}}) along the suction side of the SD7003 airfoil (geometry depicted by gray line): from left to right are zone $ab$, zone $cd$ and zone $ef$. There are 32 positions on the suction side of the airfoil.}
        \label{fig:CNN_DNS.pdf}
\end{figure*}

We report the RANS profiles with the CNN model based corrections, as compared to the baseline (uncorrected) RANS and the DNS (ground truth/high fidelity) profiles in Figs. \ref{fig:CNN_DNS.pdf} (a) and (b), respectively. The normalized turbulence kinetic energy profiles are equally spaced for the $ab$ and $cd$ zone with $x/c = 0.01$, and a spacing of $x/c = 0.02$ is used for the $ef$ zone. These normalized turbulence kinetic energy profiles are more densely packed for the $ab$ and $cd$ zones, due to the separattion and re-attachment in this region. 
From Figs. \ref{fig:CNN_DNS.pdf} (a) and (b), the CNN corrected RANS profiles exhibit a similar trend as that for the ground truth dataset and the baseline RANS predictions, as both profiles show a gradual increase in the $ab$ and $cd$ zone. Then a reduction of the profile is observed further downstream in the $ef$ zone. 

Further, in Fig. \ref{fig:CNN_DNS.pdf} (a), CNN corrected RANS profiles in general increase in magnitude as the flow moves further downstream, which is qualitatively similar to the ground truth profiles. It should be noted that the CNN-corrected RANS profiles increase in a somewhat larger magnitude than that for the ground truth right at the outset of the $ab$ zone. The discrepancy gradually reduces as the flow moves further downstream, which indicates that better accuracy of our CNN model is yielded further downstream. This behavior becomes more clear for the $cd$ and $ef$ zone. In the region where the end of the $cd$ zone meets the beginning of the $ef$ zone, the ground truth profiles are clustered due to the complex flow feature of the reattachment \cite{chu2022quantification}. Our CNN model successfully captures this clustering behavior, as shown in Fig. \ref{fig:CNN_DNS.pdf} (a), albeit with a lesser degree of intensity than the ground truth. In the $ef$ zone, our CNN model gives overall accurate predictions for the normalized turbulence kinetic energy profiles, i.e., the CNN corrected RANS profiles and the ground truth profiles are almost identical. Nevertheless, in Fig. \ref{fig:CNN_DNS.pdf} (b), the comparison shows an overall relatively large discrepancy between the CNN-corrected RANS profiles and the baseline profiles across every zone. Additionally, the baseline profiles do not show a discernible clustering behavior as observed in the CNN-corrected RANS profiles, which becomes apparent in the ground truth DNS profiles. This suggests the superior performance of the current CNN-based correction function for accurately constructing $k^{*}$.

 \bibliographystyle{elsarticle-num} 
 \bibliography{cas-refs}





\end{document}